# A possible mechanism for self coordination of bi-directional traffic across nuclear pores

Ruti Kapon[1], Alon Topchik[1], David Mukamel[2] and Ziv Reich[1]

Depts. of [1]Biological Chemistry and [2]Physics of Complex Systems, Weizmann Institute of Science, Rehovot 76100, Israel.
Email: ruti.kapon@weizmann.ac.il and Ziv Reich, ziv.reich@weizmann.ac.il

**Abstract.** Nuclear pore complexes are constantly confronted by large fluxes of macromolecules and macromolecular complexes that need to get into and out of the nucleus. Such bi-directional traffic occurring in a narrow channel can easily lead to jamming. How then is passage between the nucleus and cytoplasm maintained under the varying conditions that arise during the lifetime of the cell? Here, we address this question using computer simulations in which the behaviour of the ensemble of transporting cargoes is analyzed under different conditions. We suggest that traffic can exist in two distinct modes, depending on concentration of cargoes and dissociation rates of the transport receptor-cargo complexes from the pores. In one mode, which prevails when dissociation is quick and cargo concentration is low, transport in either direction proceeds uninterrupted by the other direction. The result is that overall-traffic-direction fluctuates rapidly and unsystematically between import and export. Remarkably, when cargo concentrations are high and dissociation is slow, another mode takes over in which traffic proceeds in one direction for a certain extent of time, after which it flips direction for another period. The switch between this, more regulated, mode of transport and the other, quickly fluctuating state, does not require an active gating mechanism but rather occurs spontaneously through the dynamics of the transported particles themselves. The determining factor for the behaviour of traffic is found to be the exit rate from the pore channel, which is directly related to the activity of the Ran system that controls the loading and release of cargo in the appropriate cellular compartment.

1.  **Introduction**

In interphase eukaryotic cells, exchange of material and information between nucleus and cytoplasm occurs through Nuclear Pore Complexes (NPC). Embedded in the nuclear envelope (NE), these unique multi-protein assemblies select and support the traffic of a myriad of substrates having various biochemical characteristics and roles as well as sizes. The molecules that pass through this conduit can reach up to 25 nm for RNP's [1] and 39 nm for virus capsids [2]. The type of molecules that make use of the NPC range from ions and metabolites to proteins, RNAs and in certain cases, DNA. Yet, NPCs are not differentiated into those that handle a particular type of substrate, nor into those that handle a particular transport direction. Furthermore, there is no known means of communication between the two edges of the NPC that would allow entry to be blocked from one side while the channel is transporting in the other direction. This raises the question of how two-way macromolecular traffic is synchronized under varying conditions to allow cell homeostasis.

The structure and the biochemistry of both the soluble phase of the nucleocytoplasmic transport machinery and of the NPC itself have been under intense study. In the next few sections, we limit our description to facts necessary for development of a simple model to study two-way macromolecular traffic. For a more detailed description please see excellent reviews by [3-11]. As revealed by cryo-electron microscopy [12-19] and more recently by a novel computational scheme [20], the NPC is a cylindrically symmetric protein conduit embedded in the fusion between the two lipid bilayers that make up the nuclear envelope. Conceptually, it can be divided into three regions: cytoplasmic, central and nucleoplasmic. The central part, which spans the nuclear envelope, measures 30 nm and 40 nm in length for yeast and metozoa respectively. It is partially occupied by filamentous proteins which contain arrays of hydrophobic peptide repeats of the form FG, GLFG or FXFG. It is widely believed that the hydrophobic repeats contribute to the selectivity of the NPC, although the mechanism is still under debate [21-23]. One model holds that the FG repeats form a sieve-like mesh [24,25] whereas another view holds that they form a polymer brush [26,27]. Along the direction of transport the NPC has an hourglass shape with the two portals of the central region measuring approximately 100 nm – 120 nm and the waist located in the centre reaching a diameter of 40 nm – 50 nm [14,15,19]. The cytoplasmic region of the NPC consists of 8 kinked protrusions whose length is approximately 35nm – 50 nm. Another 8 protrusions emanate from the nuclear side and are joined together by a distal ring, forming a structure referred to as the nuclear basket. The length of this region is 60 nm -75 nm and the diameter of the ring at its termination is 40 nm – 55 nm. All in all, the route which a substrate must travel to cross between nucleus and cytoplasm is approximately 120 nm – 180 nm.

For molecules smaller than 40 kDa (~ 9 nm in diameter), the NPC is just a conduit through which they diffuse. However, towards most proteins and towards RNP's, whose size is above this limit, the NPC exhibits selectivity. In order to pass, macromolecules must carry special signals termed Nuclear Localization Signal (NLS) and Nuclear Export Signal (NES), which allow molecules to attach to dedicated transport receptors. Once the journey is complete, receptors are detached from the cargo. On a macroscopic scale directionality is governed by a gradient in the nucleotide-bound state of the small GTPase Ran, which coordinates these attachment and detachment processes [28-30]. Ran is found predominantly in its GTP form in the nucleus and in its GDP form in the cytoplasm; the different affinities of receptors to cargoes in the presence of a particular form of Ran lead to directionality of transport on a macroscopic scale. Maintaining the Ran GTP/GDP gradient across the nuclear envelope requires energy which is obtained from hydrolysis of the Ran-bound GTP in the cytoplasm.

Notably, each NPC must support transport of macromolecules or macromolecular complexes both into and out of the nucleus, a task that is achieved with impressive efficiency at rates that can reach a few hundred translocations per NPC per second [31]. The details of transport of a single molecule through the NPC are just beginning to emerge, but evidence from electron microscopy indicates that even for small molecules, at any given time a single file of cargoes is found in the channel [32]. This single file need not necessarily be in a straight line, and routes found in different micrographs do not overlap completely. The superposition of the locations of many transporting cargoes as followed by both EM studies [15,33] and single molecule fluorescence studies [34-36] suggest that there are many possible trajectories in the NPC. Therefore, the sum of paths followed by many molecules covers a large volume of the central channel and the paths of molecules that are transporting in both directions are likely to intersect from time to time. In addition to creating a local interruption, through competition for space, this could slow down traffic in the channel as a whole. This situation can arise even for molecules which are significantly smaller than the radius of the central channel.. There is no known mechanism for separating import from export traffic in an individual NPC *i.e.*, the pores are not differentiated into dedicated import and export conduits and there is no evidence for an active gating mechanism. How then is the bidirectional traffic of particles, some of which approach the size of the pore opening, coordinated?

To address this question, we employ a non-equilibrium statistical mechanical model that incorporates the fundamental traits of receptor-mediated transport through the NPC. We do not aim at introducing a detailed representation of the molecular processes involved in the transport through NPC's. Rather, we apply the simplest possible model which incorporates only hard core interactions and directionality of the moving molecules, in order to explore the resulting transport modes. The hope, based on experience in many studies of non-equilibrium systems, is that such simple models capture some generic features of the transport dynamics, which are present in the detailed and realistic models.

**Model**

In our examination we study the behaviour of the ensemble of molecules traversing the pore by taking into account the combined effect of a multitude of particles engaged concomitantly in both import and export activities. Ideally, one would write the equations of motion for particles in the proper geometry and use the rules of statistical mechanics to derive their collective behaviour. However, since the system is open and out of equilibrium, this approach must be replaced with a non-equilibrium methodology. Consistent with the EM data [32], we consider motion through the channel as occurring along a single coordinate. It is also assumed that the only interaction between cargoes is through exclusion.

Our treatment is based on a model first developed to treat 1-D driven diffusive systems by Evans et al [37-39]. This model incorporates directionality in the system at the microscopic level. Namely, in the model, a given cargo has a preference to move in its designated macroscopic direction. Previous studies of NPC transport have used either directional or non-directional motion of the cargo in the channel (40-43). Experimentally, the degree of directionality of transport at the microscopic level is still unresolved. Our treatment assumes directionality and therefore the results presented here can be used to test this assumption.

While our approach is not based on the intricate details and interactions which occur during the transport of a single macromolecular complex through the NPC channel, the simplifying assumptions agree well with the known features of NPC transport and the model provides important insights into the way traffic through the NPC may be regulated. The main idea of our approach is to introduce the simplest possible model which exhibits the key important features of transport through the NPC. Accordingly, motion of complexes in the NPC channel is described in a coarse-grained fashion allowing one to discretize space and represent the motion of the particles by a set of hops between adjacent sites on the resulting lattice. The sites in the lattice signify that a complex travelling in the channel takes up space and therefore interferes with the passage of other macromolecular complexes. They do not correspond to real binding sites with which transport receptors may or may not interact within the channel. The dynamical behaviour of the large ensemble of molecules travelling in both directions that ensues from this simple model is studied using computer simulations.

It is assumed in the model that the NPC can support particles moving simultaneously in both directions. Owing to the macroscopically driven nature of receptor-mediated transport, originating from GTP hydrolysis concomitant with the establishment of the RanGTP/RanGDP gradient, we assume that once a complex starts its passage through the NPC channel in a given direction, the probability that it will reverse its direction is zero. We note however, that we used this total asymmetry merely to simplify the description of the model and that the results of the simulation remain qualitatively unchanged (data not shown) if cargoes are allowed to reverse, as long as there exists some degree of preference for the particle types to travel in a given direction. In view of that, macromolecular complexes passing through the system are divided into two types according to their preferred transport direction: imported particles (IPs) travel from the cytoplasm to the nucleus (left to right in Fig. 1) and exported particles (EPs) travel from the nucleus to the cytoplasm (right to left in Fig. 1).

The sites of the lattice represent a volume taken up by a macromolecular complex when it is in the channel. Thus, each discrete site in the lattice can be occupied by either an IP or an EP or no particle at all (0), but not by more than one particle at a time. While a particle occupying a particular position in the lattice prevents other particles from inhabiting the same position, particles can squeeze past each other momentarily when hopping from site to site. There is no interaction between the transporting particles other than through this exclusion process.

We define the dynamics in the model by six probabilities. Particles may be introduced or removed from the channel at the two ends; cytoplasmic and nuclear. IPs are introduced from the cytoplasm (left in Fig. 1) with a probability $\alpha_{import}$, and removed at the nuclear end (right) with a probability $\beta_{import}$. EP's are introduced from the nuclear side with a probability $\alpha_{export}$ and removed from the cytoplasmic side with a probability $\beta_{export}$. Thus all interactions and parameters which affect the onset of transport are incorporated into the parameter, $\alpha$ which is given in terms of the probability per unit time of entering the channel. These factors include, but are not limited to, the local concentrations of cargoes, transport receptors and RanGTP (for export) around the pores, the association rates between them, the availability of free sites on the peripheral structures that serve for docking of the cargo-transport receptor complexes prior to translocation, and the rate constants of these interactions. Likewise, the probability of exit, $\beta$, integrates all dissociation-related interactions. $\beta$ thus depends mainly on the interactions between RanGTP and import complexes in the nucleus and on the binding of the Ran-associated effectors RanBP1 or RanBP2, and subsequently of RanGAP to Ran-containing export complexes in the cytoplasm, and can thus be modulated. Within the channel, particles may move in their designated direction with a hopping probability *r*.

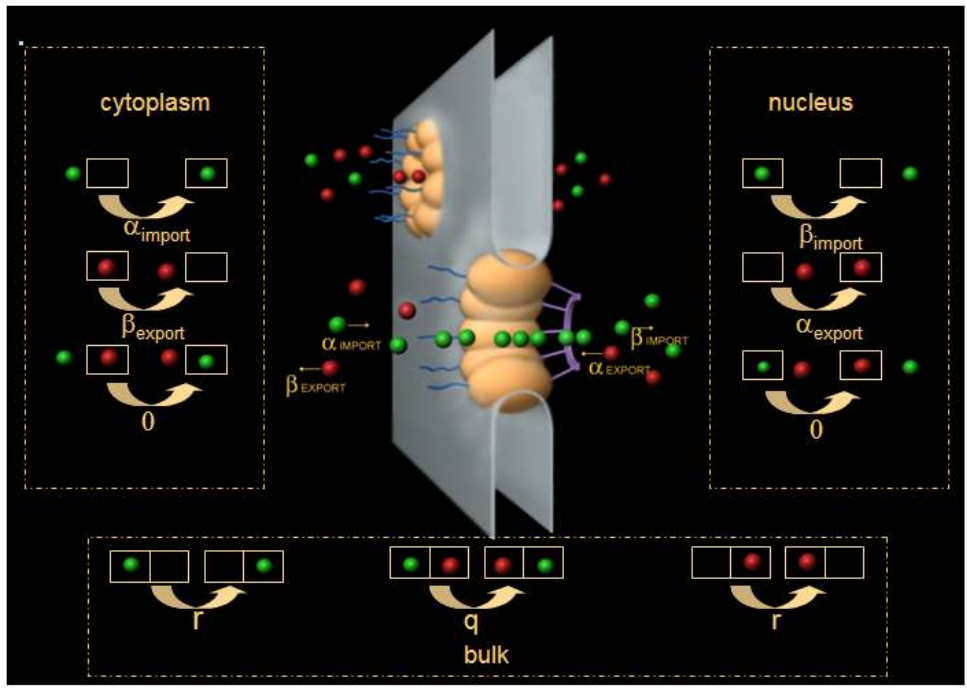

**Figure 1: Schematic of the model**. A cartoon of the nuclear pore complex allowing transport of materials between the cytoplasm, on the left, and the nucleus on the right. Import particles (IP) are shown in green and export particles (EP) are shown in red. Transit between cell compartments is described as a three-stage process involving entrance into the NPC, hopping between sites in the channel and exit into the appropriate compartment. $\alpha_{import}$ and $\alpha_{export}$ are the probability for an IP to enter from the cytoplasm and for and EP to enter from the nucleus respectively. Likewise, $\beta_{import}$ and $\beta_{export}$ are the exit probabilities for IP's and for EP's respectively. Within the channel, the probability of hopping to an empty site is given by *r* and the probability for exchange with an opposite particle is given by *q*. Throughout this paper *r=q=1*. Also shown are the rules for evolving from one generation of the simulation to the next.

The simulation proceeds as follows: Starting from an arbitrary distribution of particles within the channel, where *N* is the length of the channel, we allow the system to evolve according to the above dynamics. At each Monte-Carlo (MC) step, two adjacent sites are chosen at random and the rules above are applied to determine whether the particles have moved between sites and what the new occupation of the sites is. In order to simulate continuous movement, we take each time step to consist of *N* such MC steps. On average, every site in the channel is sampled once during each time step. Each point in the graphs presented in this work represents the average of 1000 time steps. The current in each direction is calculated by counting the number of particles that exit the channel in the appropriate compartment per unit time.

The length of the translocation path is of the order of 100 nm [14], which allows for a single file of 10 particles with an average diameter of 10 nm. Experiments, in which import of fluorescent protein markers through the NPC was followed at the single molecule level, revealed that the channel can be occupied by approximately 6-15 substrate molecules [34-36]. Because the length of the channel in the simulations is described in terms of the number of available sites, we take the length of the channel, $N$, to be 10. Complexes of different sizes may be treated by adjusting the number of sites, $N$. This variation, however, does not have a qualitative effect on the results.

In the following sections we show that the entry and exit rates have a profound effect not only on the pace of particle delivery into the proper compartment but also on the behaviour of the channel as a transport machine. In fact, modification of these two factors, which can take place through the soluble members of the transport machinery, may switch traffic between various modes of transport thus avoiding traffic jams without invoking active gating mechanisms.

## 2. Results and Discussion

To gain insight we start our treatment with the simplest situation and gradually refine it to what we believe describes the NPC more realistically.

We first discuss transport when import and export kinetics are identical. The probability for entrance into the channel is taken to be the same from both compartments: $0 < \alpha_{import} = \alpha_{export} \equiv \alpha \leq 1$, and the probability for exit of imported complexes from the NPC into the nucleus is taken to be the same as the probability for exit of exported complexes into the cytoplasm: $0 < \beta_{import} = \beta_{export} \equiv \beta \leq 1$. In this, as well as in all simulations in this work, the probability for hopping into the next site inside the channel is the same for import as it is for export and is set to 1. As a first step we will examine the effect of exit rate on the behaviour of traffic. Fig. 2(a) and 2(b) show the net current through the NPC, $J_{import} - J_{export}$, which reflects transport direction as well as magnitude, as a function of time for high (2a) and low (2b) exit rates. When cargoes can be released fast enough from the channel, the density of cargo inside the channel is low. For this case, the current is determined by the feeding rate and rapidly fluctuates between import and export, in what appears to be a random manner, as shown in Fig. 2(a).

We now turn to examine the consequences of lowering the exit rate on traffic through the channel. As discussed above, exit of the transport receptor-cargo complexes from the NPC does not occur spontaneously, but requires interactions with the Ran system [44,45]. This may lead to a delay of the complex at the channel exit and consequently to a sort of traffic jam [24]. This assumption is corroborated by results obtained from single-molecule measurements, which found exit from the central pore channel to be the rate-limiting step in NPC-mediated transport [34]. In the context of our model, this situation is described by large input rates, $\alpha$, compared with the exit rate, $\beta$. It has been shown [37] that there exists a critical exit rate, $\beta_c(\alpha)$, such that for $\beta < \beta_c(\alpha)$ the density of particles in the channel is high and the system displays spontaneous symmetry breaking (SSB) as depicted in Fig. 2(b). In this mode, the channel exists in either of two states: one in which it is predominantly occupied by import complexes and overall traffic is towards the nucleus, and the other when it is occupied by export complexes and traffic is towards the cytoplasm. Over time, transport switches spontaneously between these two states. Thus, at a given time, the overall current is predominantly in one direction only and the channel may be viewed as being dedicated to either import or to export; interference between the two processes inside the channel is largely avoided. Unidirectional transport persists for a time $\tau$, after which the channel spontaneously switches to the other direction. To understand the switching mechanism, let us assume that at some point in time the channel is in the importing state. In this case, the channel is almost exclusively occupied by import cargos that enter the system from the cytoplasmic side and leave at the nuclear side. When the exit of the channel is occupied by an IP, and $\beta < \beta_c$, the probability for an EP to enter the channel is very low , thus preserving the importing state of the channel. However, there is some probability for an empty site to form at the nuclear face of the NPC, allowing an EP to enter the channel. This particle will then move to the cytoplasmic end of the channel and stay there for some time due to the low exit rate. Because of a local stochastic fluctuation, a group of EPs may therefore temporarily form at the cytoplasmic side. When this happens, the newly incorporated EPs will slow down the entrance of imported cargo, and a cluster of empty sites, which partitions between the EPs and the IPs will form. A typical configuration would have the form (EP EP 0 0 IP IP IP IP IP IP). This state will usually be short lived and the system will relax back to its importing state. However, the longer this state persists, the larger the unfilled partition between cargoes becomes, until, after a long enough time, the system will be completely empty, allowing for the possibility of its being filled with either type of particle. If it becomes filled with EPs, the NPC may switch to an exporting state until another switch occurs. The time a channel is dedicated to a given direction, or flip-time, has been shown to increase exponentially fast with the number of available positions in the channel [37-39].

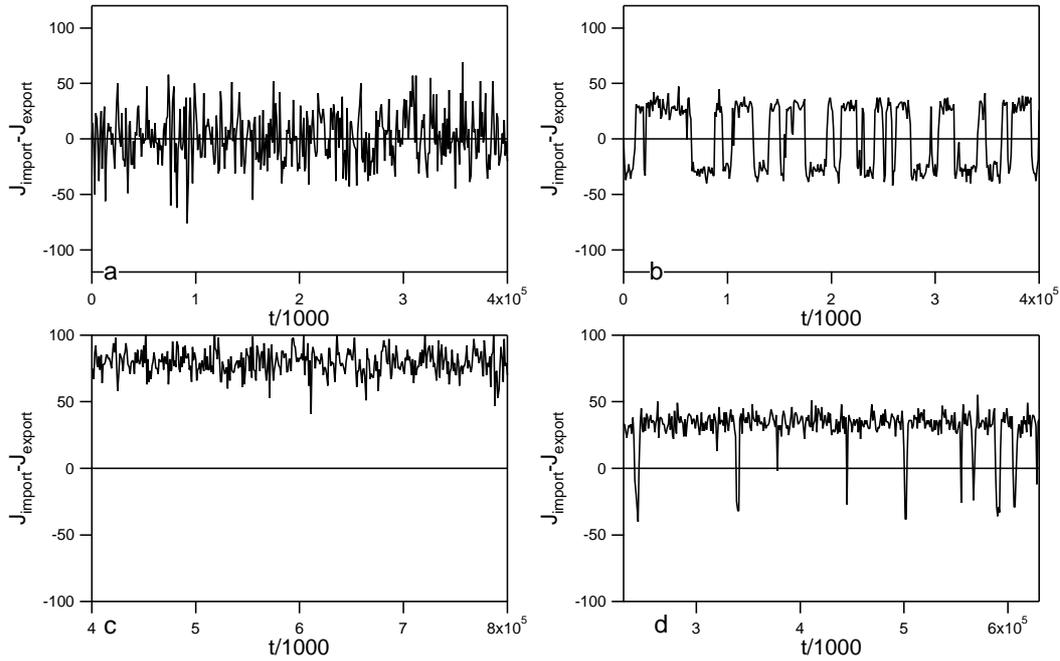

**Figure 2**: **Alternative modes of transport.** The calculated net import current, $J_{import}-J_{export}$, is plotted as a function of time. Positive values indicate that the overall current is in the direction of the nucleus. The different behaviors of the channel for the different values of entrance-exit-hopping rates are clearly seen. *(a)* Symmetric case for large $\beta$. Simulation parameters are; $\alpha < r$, $\alpha_{import}=\alpha_{export}=0.6$, $\beta_{import}=\beta_{export}=1$. The current is seen to fluctuate rapidly between import and export. *(b)* Symmetric case with a small $\beta$. $\beta << \alpha < r$, $\alpha_{import}=\alpha_{export}=0.6$, $\beta_{import}=\beta_{export}=0.08$. The channel flips between two phases; import and export with a characteristic time $\tau$. *(c)* Asymmetric case for which $\alpha_{import} \neq \alpha_{export}$ and $\beta$ is large. Simulation parameters are: $\alpha_{import}=0.6$, $\alpha_{export}=0.15$, $\beta_{import}=\beta_{export}=0.2$. The current is seen to fluctuate but remains overall in the nuclear direction. *(d)* Asymmetric case for which $\alpha_{import} \neq \alpha_{export}$ and $\beta$ is small. Simulation parameters are; $\beta << \alpha < r$, $\alpha_{import}=0.6$, $\alpha_{export}=0.15$, $\beta_{import}=\beta_{export}=0.08$. The channel can exist in two phases, importing or exporting, but it spends a considerably longer time in the importing direction to accommodate the higher load presented to it at the cytoplasmic side. For all figures, time is measured in units of Monte Carlo sweeps.

Hence, depending on the exit rate, two distinct behaviours may ensue. When the exit rate is high the channel allows traffic direction to fluctuate rapidly. On the other hand, when the exit rate is low, transport may become intermittently unidirectional. We note that exit rate in NPC transport may depend on the availability of soluble members of the transport machinery and hence may vary under varying loads. Let us look for example at dissociation in a particular region in the nucleus. When the number of molecules reaching the nuclear basket is low, local RanGTP concentrations are high enough to dissociate them immediately, thus exit rate is high. On the other hand, if the load is high there may not be enough RanGTP to dissociate cargoes from the channel, leading to the second mode of transport. A parallel situation will occur in the cytoplasm, this time the dissociation rate being dependent on the availability of free RanBP1/2, and RanGAP.

Let us now consider a more complex situation where the amount of material to be transported in the two directions is not the same, leading to an inherent asymmetry in the system. Imagine that, due to an external stimulation, an excess amount of import material is present in the cytoplasm. In this case the probability of an IP entering from the cytoplasmic side is higher than that of an EP entering from the nuclear side, $\alpha_{import} > \alpha_{export}$, and we expect the channel to support a larger overall current in the nuclear direction. As in the symmetric case, however, the mode by which this occurs depends on the exit rate. If material can be released from the channel fast enough, then the density of cargoes in the channel is low and they do not interfere with each other. The asymmetry in the number of cargoes can be balanced simply by maintaining a higher flux in the import direction, as shown in Fig. 2(c). If, on the other hand, $\beta$ is smaller than $\beta_c$, the density of cargoes is high and interference may occur. In this case, the channel will display the intermittent unidirectional behaviour described previously. In order to allow for an overall larger flux in the nuclear direction, the channel will become dedicated to import for much longer periods of time than to export (Fig. 2(d)). Thus, for both the symmetric and nonsymmetric case, the dynamic behaviour, and not only the flux of the system, is determined by the exit rate from the channel.

So far, we have considered the behaviour of the system when all parameters are held constant during the simulation. Let us now examine the case in which the number of molecules destined for transport changes over time. The concentration of a particular import or export cargo may fluctuate in response to cellular demands, but since there are many types of molecules that make use of the NPC this number is never zero. As an example however, we look at only one particular cargo type and therefore treat a somewhat synthetic case in which the number of cargoes to be transported varies significantly in time. Initially, there is a large number of molecules to be transported and this number decreases until all molecules have reached their destination compartment. We assume that the probability for a molecule to enter the channel has a simple dependency on the number of molecules around the channel entrance. We therefore take $\alpha$ to be a product of the number of particles that remain to be transported and the probability of a single molecule entering the channel $\alpha_{import,export} = \alpha_{initial}\, n(t)/n(0)$, where $n(t)$ is the time-dependent number of particles in the compartment. Each time a cargo enters the channel, the number of particles that remain to be transported is reduced by 1, and $\alpha$ changes accordingly, until eventually, all particles have reached their destination. These simulations were performed for the case in which there is more material to be imported than exported, *IP > EP*, and the initial entry rate to the channel is higher than the exit rate, $\alpha_{initial} > \beta$ with $\beta<\beta_c$.

The first obvious feature, shown in Fig. 3(a), is that the time the channel spends in the import state is significantly longer than that spent in the export state. This is because there are more particles to be imported than exported. As time goes by, the number of particles that still remain to be transported decreases on both sides; thus both $\alpha_{import}$ and $\alpha_{export}$ decrease with time until eventually $\alpha(t) < \beta \leq \beta_c$ and transport direction starts to fluctuate rapidly. Panel *B* of Fig. 3 shows the number of cargoes that have accumulated in the destination compartment as a function of time. Note that because the channel is dedicated to a particular direction of transport at each given time, the graphs include steps in which particles accumulate in the nucleus, while the number of particles accumulating in the cytoplasm remains roughly constant and vice versa.

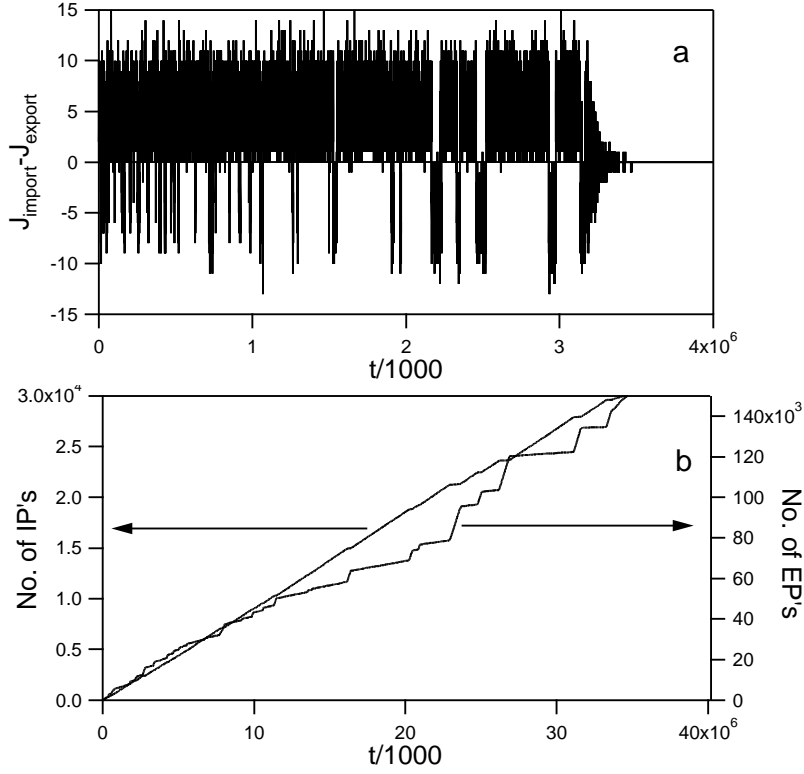

**Figure 3: Transport dynamics for a time-dependent $\alpha$.** In this simulation, the $\alpha$'s depend on the number of particles in the reservoirs and are therefore time dependent; $\alpha_{import}=1 \cdot IP(t)/IP(0)$, $\alpha_{export}=1 \cdot EP(t)/IP(0)$, $\beta_{import}=\beta_{export}=0.01$. The initial number of particles is $EP(0)=3 \cdot 10^4$ and $IP(0)=1.5 \cdot 10^5$. *(a)* Time evolution of the current difference. Current direction flips but the time period when the channel is dedicated to import is significantly longer, since $\alpha_{import} > \alpha_{export}$. This advantage is clear and sustained until the number of particles on both sides becomes equal ($t \sim 33 \cdot 10^6$) and traffic direction flips rapidly. *(b)* The number of particles accumulated in the destination compartments. The curves describing particle accumulation contain steps that correspond to uni-directional traffic.

Next, we wish to examine a situation where the exit rates from the channel are not equal in both compartments. To isolate the effect of exit rates, we go back to the case in which the initial number of particles to be transported is the same in both compartments. These numbers are updated self-consistently by subtracting from them the number of particles that have already been delivered to the appropriate compartment. This allows us to follow the temporal behaviour of the system until all particles have reached their destination. We set the exit rate from the channel into the cytoplasm higher than into the nucleus. Thus the dissociation of exported particles from the NPC is faster than that of imported particles. The resulting currents in both directions are shown in Fig. 4. Surprisingly, we find that initially the NPC becomes entirely dedicated to import. Only after almost all of the complexes to be imported have reached their destination, does the channel flip direction, and material is exported through the NPC. Notice, however, that the import current is significantly lower than the export current. Thus, for asymmetric dissociation rates, one may describe transport as proceeding in two stages. The first is a prolonged stage which consists of a low current in the direction in which $\beta$ is low. The second stage is shorter but involves higher currents. The low exit rate in the cytoplasm, which leads to a low export flux, is compensated for by the proportionally longer time the channel spends exporting. This suggests that for cases where there is a temporary shortage of dissociation factors in either compartment and $\beta$ is reduced, a stochastic mechanism exists which will modulate traffic direction to compensate for this reduction. Conversely a local increase in dissociation rate in a particular compartment will empty the channel of all cargoes moving towards this compartment, leaving it free for cargoes transporting in the other direction. This offers a means for the soluble phase of the system (*i.e.,* the Ran system, transport receptors and the cargoes themselves) to adjust traffic in response to changes in cellular demands without having to transfer information across the NPC to the other compartment.

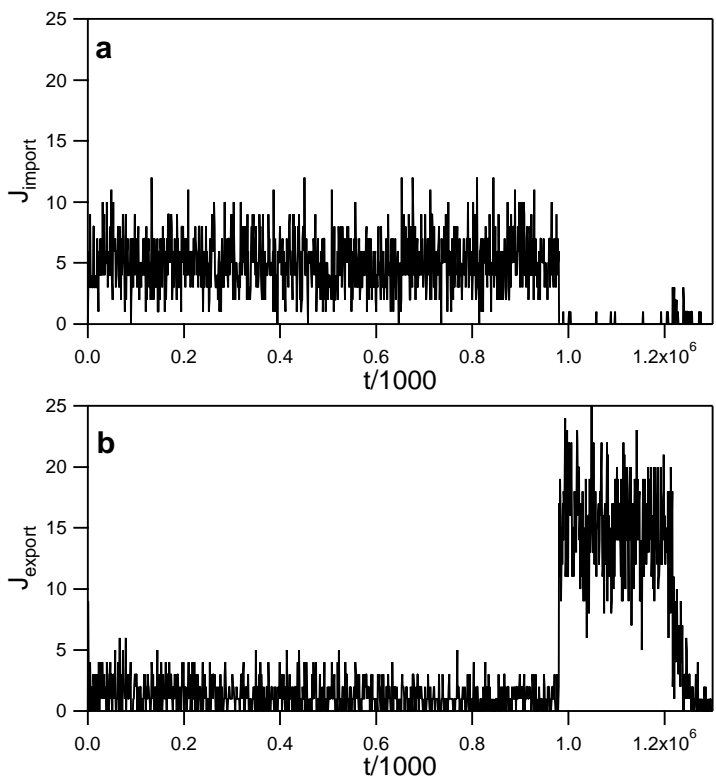

**Figure 4: $\beta$ determines traffic direction.** *(a)* Import current. *(b)* Export current when cytoplasmic release is faster than nuclear release. Simulation parameters are; $\alpha_{import} = 1 \cdot IP(t)/IP(0)$, $\alpha_{export} = 1 \cdot EP(t)/EP(0)$, $\beta_{import}=0.01$, $\beta_{export}=0.03$. The initial number of particles is $1 \cdot 10^5$ on both sides. For the low values of $\beta$ used in the simulation, exit from the channel is the rate limiting parameter of transport. The asymmetry in $\beta$ results in prolonged stage characterized by a low flux in the direction for which dissociation is slow. This is followed by a short, high flux in the opposite direction where complex release is fast. This behaviour represents a trade off between the time the channel is dedicated to a particular direction and the corresponding flux.

3. **Conclusion and Outlook**

We have examined bidirectional traffic through the Nuclear Pore Complex in the context of the collective behaviour exhibited by transported material. We show that coordination of import and export through the NPC can readily be rationalized if the behaviour of the ensemble of transported macromolecules is considered rather than that of each individual molecule separately.

We demonstrate that stochastic processes can provide the nucleocytoplasmic transport machinery a natural, self-adjustable means to cope with variations in incoming fluxes imposed by changing cellular demands, without requiring an elaborate 'decision making' process by the NPC. Rather than invoking an active gating mechanism, for which there is no experimental evidence, our treatment shows that synchronization of import and export lies in the *dynamic evolution* of the transport process itself. We find that transport dynamics is determined to a large extent by the rate of release of cargo from the channel. The statistics of the process, together with the simple set of rules for the dynamics, dictates that, when possible, the direction of transport will fluctuate quickly, but when necessary, the channel will become intermittently unidirectional. For example, a temporary reduction in the ability to release cargo in the nucleus due to a local shortage of RanGTP is compensated for by a prolonged period in which the NPC is dedicated to import. Thus, the model directly relates the activity of the soluble members of NPC machinery to transport rate and directionality through the parameters $\alpha$ and $\beta$ where $\alpha$ can be shown to determine the number of particles that are transported in each unidirectional batch and $\beta$ the time duration of the batch.

In this picture, the role of the Ran system in determining directionality becomes even more apparent in that it directly affects the mode of traffic.

We note that the behaviour predicted by our analyses rely on the existence of some degree of microscopic directionality in nucleo-cytoplamic transport. The microscopic details of this transport process are still under debate. Perhaps the dependence of our predictions on directionality could be used to experimentally test for the existence of microscopic drive.

**Acknowledgements**

We thank Guy Ziv, Eitan Bibi and Bracha Naim for helpful discussions and comments. The work was supported by the Minerva Foundation with funding from the Federal German Ministry for Education and